# A Supersonic Ping Pong Gun

Mark French
Craig Zehrung
Jim Stratton

Department of Mechanical Engineering Technology
Purdue University

**Introduction**
The Ping-Pong gun or Ping-Pong bazooka [1-3] has been a popular and compelling tool for demonstrating acceleration and the power of pressure differentials in air. It has even proven to be a challenging problem for computational fluid dynamics [4] since it encompasses a range of aerodynamic phenomena that are particularly difficult to model. However, the ball emerging from the barrel is necessarily traveling at subsonic speed. A simple modification to the basic design allows the device to launch the ball supersonically.

Consider the Ping-Pong gun as a small wind tunnel. A type of wind tunnel sometimes called a blow down tunnel [5] propels air through the test section by venting a pressure chamber into it. The primary use of blow down tunnels is for supersonic testing, made possible by a convergent-divergent nozzle between the pressure tank and the test section.

Fundamental principles of compressible flow dictate that that the Mach number in the test section is dependent on the pressure ratio across the nozzle and the geometry of the nozzle itself. It is a logical extension of the original design of the Ping-Pong gun to add a small pressure chamber and a convergent-divergent nozzle as shown in Figure 1.

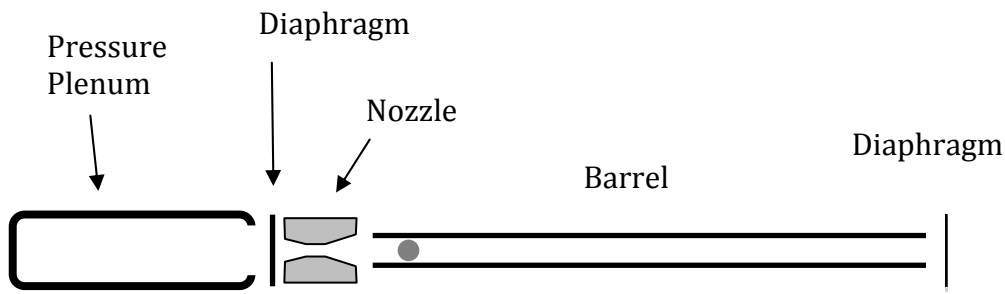

**Figure 1** – Layout of the Supersonic Ping Pong Gun

We have made an example of this device and have achieved launch speeds of 406 m/sec. Assuming the speed of sound is approximately 330 m/sec, this corresponds to a Mach number of about 1.23.

**Modifications to Standard Ping Pong Gun**

The two most significant modifications to the standard ping pong gun are the addition of a pressure plenum and a convergent divergent (deLaval) nozzle between the pressure chamber and the barrel.

The pressure chamber is comprised of 3 inch (7.62cm) diameter schedule 80 gray PVC. It is 36 inches (91.4cm) long and is able to withstand pressures up to 120psi (827kPa). One end is sealed through the use of a schedule 80 PVC cap. The other end is sealed using 'Duck Tape' brand lamination film. This film is placed around the free end of the pressure chamber and threaded into the pressure side of the nozzle. Depending on the desired pressure, the thickness of the lamination tape is either doubled or tripled in thickness. From previous testing, we found that 2-ply sheets will generally burst at a pressure differential of about 414 kPa (60 psi) and 3-ply sheets will burst at a pressure differential of around 620 kPa (90 psi).

Note that the pressure gauge we used was simply threaded into the pressure chamber. Thus, it read pressure referenced to atmospheric pressure. The actual pressure differential was one atmosphere higher than that displayed on the gauge.

It is important that the pressure chamber be able to withstand pressures much higher than those expected during use. The compressed air in the plenum contains significant energy that could be released uncontrolled if the plenum failed structurally.

The nozzle was machined from a 4.5 inch (11.4cm) diameter by 12.125 inch (30.8cm) long piece of solid PVC. Instead of machining threads into the end of the nozzle, a sch80 PVC coupler was cut down and glued into the end. The breach end of the nozzle is oversized for the nominal outer barrel diameter. In order to ensure that the barrel could be easily removed, we chose to oversize the breach end of the nozzle. In this manner, two bushings are placed around the barrel to ensure a snug fit, and to aid in removal. Also, to ensure that both the nozzle and barrel were properly sealed and able to hold a vacuum, 'Sticky Tack' was used around the interface between the nozzle and barrel. Dimensions of the nozzle are shown in Figure 2.

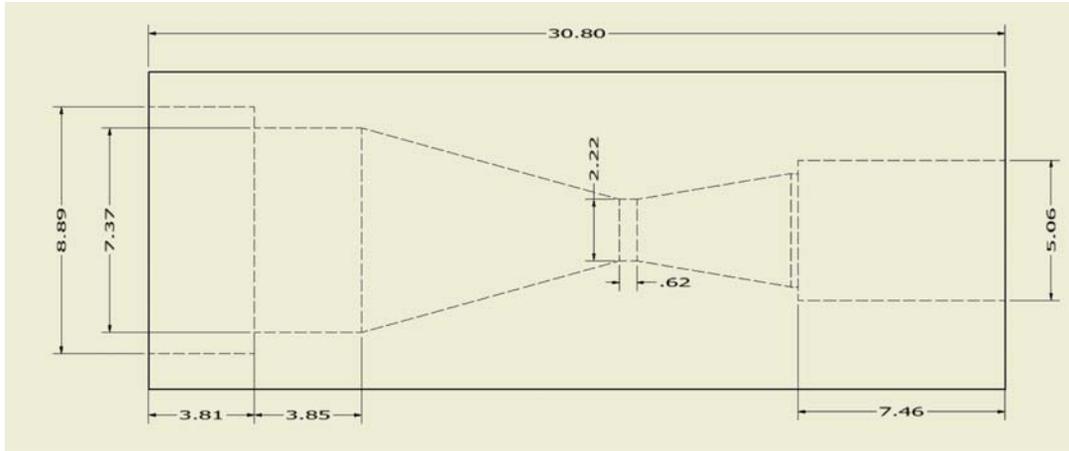

**Figure 2** – Nozzle Dimensions (in centimeters)

The barrel is comprised of 1.5 inch (3.81cm) diameter x 96 inch (243.8cm) long, clear, sch40 PVC. The overall dimensions of the device are shown in Figure 3.

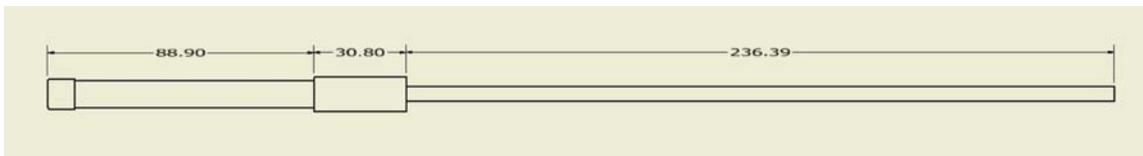

**Figure 3** – Overall Dimensions (in centimeters)

The diaphragm between the pressure plenum and the nozzle is wrapped around the threads of the pressure plenum before it is screwed into place. This provides a good mechanical grip on the diaphragm that is also airtight. This is shown in Figure 4.

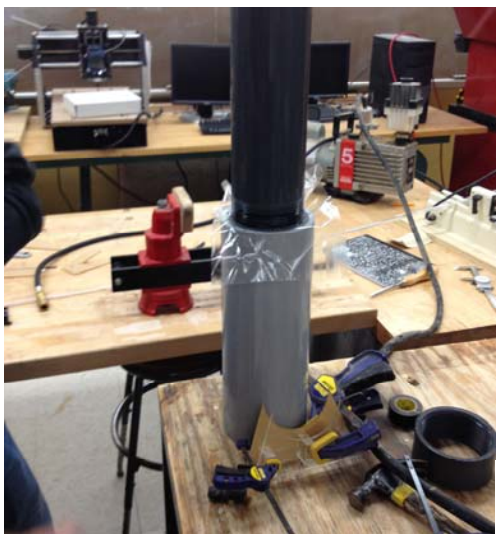

**Figure 4** – Nozzle Joined to the Pressure Plenum with Diaphragm in Place

The barrel is simply press fit into the nozzle as shown in Figure 5.  After assembly, a bead of a flexible polymer ('Sticky Tack') is pressed into the junction of the barrel and the nozzle.  This is the blue strip visible on the right side of Figure 5.

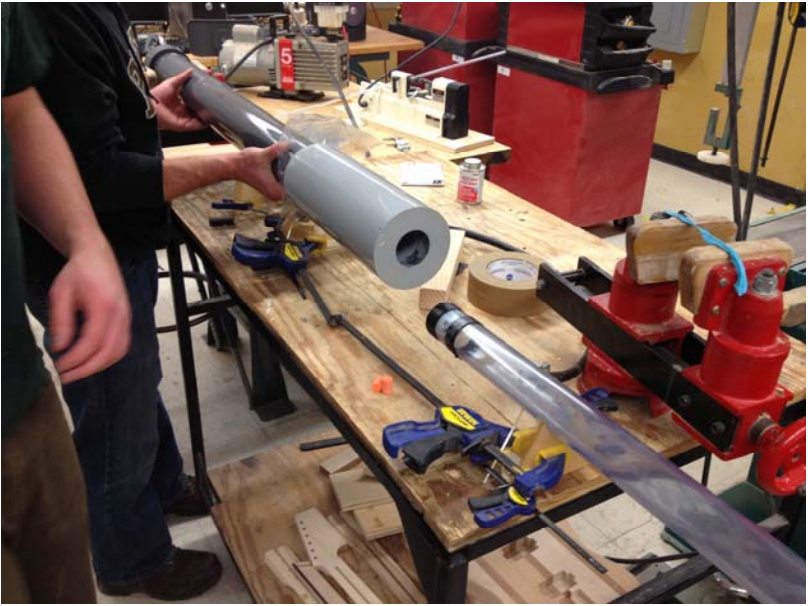

**Figure 5** – Barrel Being Fitted to the Nozzle

**Measured Performance**
We measured the velocity of the ball using a high-speed video camera set to record at 16,000 frames/second with an exposure time of 1/128,000 second.  Figure 6 shows a representative frame.  The speed of the ball was measured through the use graduated calibration grid with 0.25 inch (0.635cm) spaces.  Velocity variation between shots was relatively small, a few percent.  Video results from a typical shot gave a velocity of 406.4 m/sec, which corresponds to Mach 1.23.

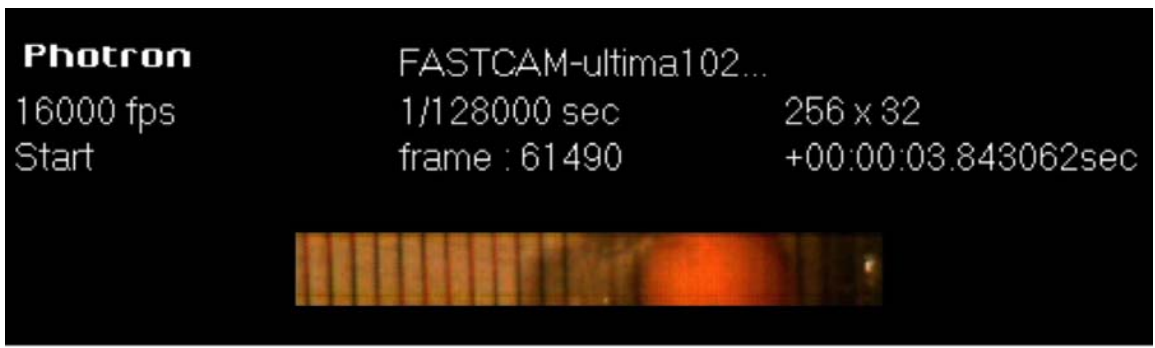

**Figure 6** – High Speed Video Frame Showing Ball in Flight

**Conclusion**

By adding a pressure chamber and a convergent divergent nozzle, a standard Ping-Pong gun has been shown to launch the ball supersonically.

There has not yet been an analysis to confirm the effect of the nozzle, so the design must still be considered as an empirical one. Further study is needed to confirm that the nozzle works as assumed. Varying the nozzle geometry may have a major effect on launch velocity.

Finally, it is important that the device be used only in a controlled environment. The increase in velocity greatly increases the kinetic energy of the ball, so care must be taken to ensure that the ball is captured after launch. Additionally, it is very important to keep all observers behind the muzzle and safe from any bounces or ricochet.